\documentstyle[prd,aps,preprint,epsfig,floats]{revtex}
\tightenlines

\def\DESepsf(#1 width #2){\epsfxsize=#2 \epsfbox{#1}}

\begin{document}

\draft

\preprint{\vbox{
\hbox{UMD-PP-05-002}
}}

\title{{\Large\bf Possible Gauge Theoretic Origin for Quark-Lepton
Complementarity}}
\author{\bf P.H. Frampton$^1$ and R.N. Mohapatra$^2$}

\address{$^1$Department of Physics and Astronomy, University of North Carolina, Chapel
Hill, NC 27599-3255.\\
$^2$ Department of Physics, University of Maryland, College Park,
MD 20742, USA}
\date{July, 2004}
\maketitle

\begin{abstract}
Similarity between the weak interaction properties of quarks and
leptons has led to suggestions that the origin of lepton mixing
angles may be related to those of quarks. In this paper, we
present a gauge model based on $SU(2)_L \times SU(2)_R\times
SU(4)_c$ group that leads to a new form for the quark lepton
complementarity which predicts the solar neutrino mixing angle in
terms of the Cabibbo angle for the case of inverted mass
hierarchy for neutrinos. We also indicate how these ideas can be
implemented in an $E_6$ inspired trinification $SU(3)_C \times
SU(3)_L \times SU(3)_R$ model, which is more closely allied to
string theory by the AdS/CFT correspondence.

\end{abstract}

\vskip1.0in
\newpage

\section{Introduction}
One of the major challenges for particle theory today is to understand
the totally different pattern of mixings among neutrinos
compared to quarks. Although many
different ways have been proposed to tackle this challenge (for recent
reviews, see \cite{af}), no final consensus has emerged on what is the
most promising way to proceed. Any new approach is therefore
desirable.

In this paper, we will focus on understanding the solar mixing
angle, $\theta_{\odot}\equiv \theta^\nu_{12}$, which is large but known
not to be maximal. Since there are leptonic symmetries that lead to
maximal solar mixing, a  possible approach to this problem is to start
in that limit and understand the deviation using new physics.
There have been many attempts in the literature to achieve
this goal\cite{many}. In this note, we pursue
a recent suggestion\cite{raidal,smirnov}  according to which the deviation
from maximality
of the solar mixing may be related to the quark mixing angle
$\theta_C\equiv \theta^{q}_{12}$. It is based on the
observation that the mixing angle responsible
for solar neutrino oscillations, $\theta_{\odot}\equiv \theta^\nu_{12}$
satisfies an interesting
complementarity relation with the corresponding angle in the quark sector
$\theta_{Cabibbo}\equiv \theta^q_{12}$ i.e. $\theta^\nu_{12}+\theta^q_{12}
\simeq \pi/4$. This equation is satisfied to within a few per cent
accuracy. While it is quite possible that this is purely accidental, it
is interesting
to pursue the possibility that there is a deep meaning behind it
and see where it leads. We explore the possibility that this relation may
be a reflection of a fundamental quark-lepton unification.

To see how quark lepton unification can possibly lead to a
complementarity relation, first note that $\theta_{\odot}$ is the
$12$ entry of the PMNS matrix defined as $U_{PMNS}=
U^{\dagger}_\ell U_\nu$ (where $U_\ell$ is the unitary matrix
operating on the left handed charged leptons that diagonalizes the
$(e,\mu,\tau)$ mass matrix and $U_\nu$ is the corresponding one
for the neutrino Majorana mass matrix) and $\theta_c$ is the
corresponding entry in the CKM matrix defined as
$V_{CKM}=U^{\dagger}_u U_d$ (where $U_{u,d}$ diagonalize the up
and down quark mass matrices). Secondly suppose that the
structure of neutrino and quark mass matrices at the high scale
are such that to a leading order, the PMNS matrix is exact
bimaximal\cite{barger1} whereas the CKM matrix is an identity
matrix and to next leading order we have the down quark and
charged lepton mass matrices are equal (or very nearly so) in the basis
where the up
quark and neutrino Dirac masses remain unchanged, then a
complementarity relation between quarks and leptons can emerge
from quark-lepton unification at high scale.

 In this paper, we make an attempt to connect the deviation of solar
mixing angle from its maximal value to the Cabibbo angle using the gauge
group $ SU(2)_L\times SU(2)_R\times SU(4)_c$
symmetry that unifies quarks and leptons\cite{ps} at high scale. We find
that if we use
the double seesaw formula for understanding small neutrino masses, then
the departure of the solar angle from maximality is determined by the
Cabibbo angle and the prediction for the solar mixing angle is consistent
with experiments within present uncertainies. This model also predicts a
``large'' value for $\theta_{13}\simeq 0.15$.

We also indicate how these considerations may be extended to the
case of a gauge group $[SU(3)]^3$ which unifies to the
gauge group $E_6$ at high scale, that has quark-lepton unification.

\section{Gauge model for quark-lepton complementarity (QLC) in the 1-2
sector}
Before proceeding to the construction of the model, we first outline
a set of conditions that are sufficient for obtaining a connection between
the $\theta_\odot$ and $\theta_C$.
 The full quark and lepton mass matrices should be written in a
form such that, we have:
\begin{eqnarray}
M_{u,d}~=~M^0_{u,d}+\delta M_{u,d}\\
M_\ell~=~M^0_\ell + \delta M_\ell\\
{\cal M}_\nu~=~U^*_{bm}{\cal M}^d_\nu U^{\dagger}_{bm}
\end{eqnarray}
where $U_{bm}$ is the bimaximal PMNS matrix given by:
\begin{eqnarray}
U_{bm}~=~\pmatrix{\frac{1}{\sqrt{2}} & \frac{1}{\sqrt{2}} &0 \cr
\frac{1}{2} & -\frac{1}{2} & \frac{1}{\sqrt{2}}\cr
\frac{1}{2} & -\frac{1}{2} & -\frac{1}{\sqrt{2}}}
\end{eqnarray}
 We require that $M^0_u~=~ tan\beta M^0_d$ so that to leading order
$V^0_{CKM}= {\bf I}$, where ${\bf I}$ is the identity matrix and
$M^0_\ell=  M^0_d$ due to quark lepton symmetry. We also choose
$\delta M_u=0$. Note also that we have chosen the $ \delta M_\ell
\approx \delta M_d$ at the scale where neutrino masses arise via
seesaw mechanism i.e. exact quark-lepton symmetry to zeroth order but
approximate quark-lepton symmetry  upto
the first order terms that generate the $12$ mixing angles. The
observed pattern in the $V_{CKM}$ then arises from the matrix
$\delta M_d$. Its similarity in form to $\delta M_\ell$ feeds this change
to the lepton mixing matrix in a way such that a form of
quark-lepton complementarity emerges in the end. The Cabibbo angle is
small in our model due to it being related to the quark mass ratios.
 We also note that $\delta M_d$
has to be different from $\delta M_\ell$ in order to fit the muon mass
after correcting for extrapolation effects from the seesaw scale to the
weak scale. The departure from maximality of solar mixing angle is
predicted in terms of Cabibbo angle as well as muon and strange quark
masses.

We first describe in detail how in a gauge theory based on the gauge group
$SU(2)_L\times SU(2)_R\times SU(4)_c$ the above conditions
can be satisfied. In a subsequent section, we will indicate how the same
considerations can be carried over to a trinification theory.

Let us start with the gauge group $SU(2)_L\times SU(2)_R\times SU(4)_c$
 with supersymmetry and an additional softly broken global symmetry
 on the theory given by
$U(1)_X$ where $X= f_1-f_2-f_3$ with $f_i$ representing the i-th
family. In the lepton sector this corresponds to
$L_e-L_\mu-L_\tau$ symmetry, which has been extensively discussed
in the literature\cite{emutau}. We make the usual assignment of
fermions to the $\Psi(2,1,4)+\Psi^c(1,2,4^*)$ as follows:
\begin{eqnarray}
\Psi~=~\pmatrix{u_1 & u_2 & u_3 & \nu\cr d_1 & d_2 & d_3 & e}_{L}
\end{eqnarray}
We add a set of three singlet fermions $S_i$ and $S^c_i$ to the theory
(i=1,2,3 stands for generation) which also have the same fermion
number as the normal matter multiplet. For the Higgs sector, we choose the
following multiplets: $\phi\equiv (2,2,0)$; $\chi\equiv (2,1,4)$
and $\chi^c\equiv (1,2,4^*)$. In a supersymmetric theory, anomaly
cancellation requires that we also have the fields:
$\overline{\chi}\equiv (2,1,4^*)$ and $\overline{\chi^c}\equiv
(1,2,4)$.

We break the gauge symmetry down to the standard model by the vevs of the
fields $<\chi^c>=<\overline{\chi^c}>=v_R$. The standard
model symmetry is then
broken down by the vev $<\phi>~=~\pmatrix{\kappa & 0\cr 0 & \kappa'}$.
We keep $<\chi^0>=<\bar{\chi}>=0$, which can be a stable vacuum if we
forbid the trilinear term ${\chi}\phi\chi^c$ by a discrete symmetry.

We can write down the Yukawa couplings of the model responsible for quark
and lepton masses as follows:
\begin{eqnarray}
{W}_Y~=~h_{ij}{\Psi}_{i}\phi\Psi^c_{j}+f_{ij}({\Psi}_{i}\bar{\chi}
S^c_j
+{\Psi}^c_{i}\bar{\chi}^c S_j)~+~\mu_{ij} S_{i}S_{j}
\end{eqnarray}
We choose the Majorana mass for the $S^c$ fields to be very heavy so that
they decouple at TeV energies. This also decouples the
${\Psi}_{i}\bar{\chi} S^c_j$ term from the Lagrangian, which we
wrote down above for completeness. Note that $U(1)_X$
symmetry restricts the form of $\mu_{ij}$ to be
\begin{eqnarray}
\mu~=~\pmatrix{0 & \mu_1 & \mu_2\cr \mu_1 & 0 & 0\cr \mu_2 & 0 & 0 }
\end{eqnarray}
After the $<\chi^c>$ and the $<\phi>$ vevs are turned on, we have the
following zeroth order forms for the various fermion mass matrices:
\begin{eqnarray}
\frac{\kappa'}{\kappa}M^0_u\equiv cot\beta M^0_u ~=~M^0_d ~=~M^0_\ell
\end{eqnarray}
where $M^0_u~=~\pmatrix{m_{11} & 0 & 0\cr 0 & m_{22} & m_{23} \cr
0 & m_{23} & m_{33}}$
Since the up and down quark masses are proportional, the CKM matrix is
the unit matrix. The leptonic mixing matrix is however not
proportional to a unit matrix as can
be seen below.

The mass matrix for $(\nu, \nu^c ,S)$ is given by:
\begin{eqnarray}
M~=~\pmatrix{0 & M^0_u & 0 \cr M^{0}_u & 0 & fv_R \cr 0 & f^Tv_R
& \mu} \label{ds}\end{eqnarray}
 where $M^0_u,fv_R$ are all
$3\times 3$ matrices whose forms are similar and the form of
$\mu$ is given above. First point to note is that the matrix
$M^0_u$ can be diagonalized without effecting the form of $\mu$.
In general the matrix $f$ can be written in the form
$\pmatrix{f_{11} & 0 & 0 \cr 0 & f_{22} & f_{23} \cr 0 & f_{23} &
f_{33}}$ due to the $U(1)_X$ symmetry.

The  matrix in Eq. (9) is in the double seesaw form\cite{valle}
which leads to small neutrino masses even with a multi-TeV scale
for the seesaw. The scale however can also be close to the GUT
scale\footnote{This possibility was pointed out by S. F. King
(private communication).}. In this note, we keep the double
seesaw scale $v_R$ near 100 TeV and include its effect
(via renormalization group) on the fermion masses through a
parameter $\Delta$.

Diagonalizing the full neutrino matrix in Eq.\ref{ds}\cite{valle},
 one obtains the light neutrino mass matrix ${\cal M}_\nu$ to be
\begin{eqnarray}
{\cal M}_\nu~=~\pmatrix{0 & m_1 & m_2 \cr m_1 & 0 & 0\cr m_2 & 0 & 0}
\end{eqnarray}
where $m_1= \frac{m_{11}m_{22}\mu_1}{f_{11}f_{22}v^2_R}$ and $m_2=
\frac{m_{11}\mu_1 m_{33}}{f_{11}\tilde{f}_{23}v^2_R}+
\frac{m_{11}\mu_2m_{33}}{f_{11}{f}_{33}v^2_R}$. Thus for
instance, if $m_{ij}\sim$ GeV and $\mu_i\sim$ GeV and $v_R\simeq $
100 TeV, we get for neutrino masses $m_i\simeq 0.1$ eV, which is
of the same order as the square root of the atmospheric neutrino
mass difference square. We could increase $v_R$ by appropriately
adjusting the scale $\mu_{ij}$. For instance, if $v_R$ is assumed
to be at the GUT scale, so would be $\mu_{ij}$. Clearly, if
$m_1\simeq m_2$ we also get maximal mixing needed for atmospheric
neutrino oscillation.

Note that solar mixing angle is maximal and in fact the PMNS
matrix at this zeroth order level is bimaximal for $m_1=m_2$ and the mass
ordering of neutrinos is the so-called inverted type. Of course
the model
 at the zeroth order level has the unpleasant feature that it predicts
vanishing solar neutrino mass
difference which however is corrected as we include higher order terms.

From the relation $M^0_u~=~tan\beta  M^0_d ~=~tan\beta  M^0_\ell$, it
may appear that, we have unacceptable relations among the masses
of quarks and leptons. Models with such a property have been
discussed before\cite{babu} and it has been shown how such
models, if supersymmetric can be phenomenologically fully viable.
These mass relations without further corrections imply that
$m^0_s/m^0_b=m_c/m_t$ and $m^0_\mu (M_c)= m^0_s (M_c)$, where $M_c$ is
the SU(4) scale. Note that $m_s$ and $m_\mu$ cannot be close to each
other at the multi-TeV scale since renormalization group
extrapolation due to strong interactions  increases $m_s$ without changing
 $m_\mu$ much. Therefore, we will use the $\delta M$ terms to get
the right $m_s/m_\mu$. We will see that we can get a sizable correction to
the maximality of solar mixing angle dictated bt the zeroth order terms.
 
We will assume that the $\delta M$ terms do not affect the third
generation fermion masses. The ratio $m_c/m_t$ then implies that
$m_b=m_\tau \sim 10 $ GeV. These masses are much higher than their
observed values.
It has however been noted that in supersymmetric theories, there
are large gluino and Wino contributions to the bottom and tau
masses, which can correct for this\cite{babu} and bring them down
to observed values. For large $tan \beta$, there are two
contributions to the quark masses and one for the charged
lepton\cite{mass}:
\begin{eqnarray}
\delta m_b~=~\frac{2\alpha_s m_{\tilde{g}}}{3\pi m^2_{\tilde{q}}}(m^0_b\mu
tan\beta + A^d_{33}m_0) + \frac{Y^2_t\mu}{16\pi^2
m^2_{\tilde{q}}}(m^0_b\mu+A^u_{33}m_0tan\beta)
\end{eqnarray}
and
\begin{eqnarray}
\delta m_\tau~=~\frac{g^2_1 m_{\tilde{B}}}{16\pi^2
m^2_{\tilde{l}}}(m^0_\tau\mu
tan\beta + A^e_{33}m_0)
\end{eqnarray}
Clearly, there are several independent parameters which one may
adjust to get the bottom and the tau lepton mass right. Note that
the quark mixings still vanish at this stage. As noted before, if we
choose a value for $m_s\simeq m_\mu$ somewhere around 100 MeV. For
 $v_R$ in the 100 TeV range, when we come down to the
weak scale, $m_s$ will increase and become bigger than $m_\mu$.
In the next section, we show how the higher order effects can fix
this problem, generate quark and lepton mixing leading to
quark-lepton complementarity as well as give the first generation
masses right.

\bigskip

\section{Inclusion of next order terms and
derivation of quark-lepton complementarity}

\bigskip
\bigskip

In order to obtain the CKM angles and the departure from bimaximal
pattern for neutrino mixings, we need to include higher
dimensional non-renormalizable terms in the theory that break
$U(1)_X$. We assume that they contribute only to the down quark
and charged lepton mass matrix in a quark-lepton symmetric way.
For this purpose, in the 422-model, we include a $SU(2)_R$
triplet Higgs field $\delta_R$ which transforms under the gauge
group as $(1,3,15)$ and a gauge singlet $\sigma(1,1,15)$, both of which
have $X=+2$. We will also include a $\delta_L$ with zero vev so
that it has no effect on the masses. This allows us to include two
terms which can lead to violation of $X$ charge after symmetry
breaking:
\begin{eqnarray}
\delta {\cal
L}~=~\frac{1}{M}[f'_{ij}{\psi}^T_{i}\phi\vec{\tau}\cdot
\vec{\delta}_R\psi^c_{j}+f''{\psi}^T_{i}\phi\sigma\psi^c_{j}]
+\frac{1}{M^2}( {\psi^T}_{1}\phi{\chi}^c
\bar{\chi^c}\psi^c_1)
\end{eqnarray}
where $i=1$ and $j=2,3$. As just noted, once $\vec{\delta}_R$ and
$\sigma$ acquire vevs, we can arrange their values and the couplings
$f'$ and $f^{''}$ so that the
contribution of these terms to the fermion mass appears
predominantly in the down sector. Secondly due to the fact that the
multiplets transform as {\bf 15} dim. representation of $SU(4)_c$, they 
will make different contribution to $\delta M_d$ and $\delta M_\ell$.
  We choose $<\delta_R>, <\sigma>\simeq <\chi^c>$ and $M\sim 10^2
(<\delta_R>, <\sigma>)$ so that  the magnitude of
these terms of the right
order of magnitude (i.e. $10^{-2}$ GeV for the first two and few
$\times 10^{-3}$ GeV for the last term) so that we generate the right
value for the Cabibbo angle. Clearly, the last term makes a small
contribution (due to two powers of $M$ in the denominators) to the
electron mass but not big enough to contribute to quark masses.
Collecting all these together and ignoring the third generation fermions
we can write the mass matrices
for the first two generation of quarks and leptons in the
presence of the higher dimensional terms at the weak scale:
\begin{eqnarray} M_\ell~=~\pmatrix{m_0 & -3\delta'\cr -3\delta' &
m^0_\mu-3\delta}\\ \nonumber
 M_d~=~\Delta\pmatrix{0 & \delta'\cr \delta' &
m^0_\mu+\delta}
\end{eqnarray}
where $m^0_\mu$ is the zeroth order contribution, $\delta$ and
$\delta'$ are the next order contributions from Eq. (13) and $\Delta$ is
renormalization effect on
the quark masses from $v_R$ down to $M_Z$. In this model $\theta_c\simeq
\frac{\delta'}{m^0_\mu+\delta}\equiv \sqrt{\frac{m_d}{m_s}}$. On
the other hand, the mixing angle for charged lepton is:
\begin{eqnarray}
\theta^\ell\simeq \frac{-3\theta_C m_s}{\Delta m_\mu} .
\end{eqnarray}
 We choose the scale $v_R$ to be low (in the multi-TeV range) and
choose for example of $m_s\sim m_\mu/2$ at
$v_R$ scale.   The value of $m_s$ increases somewhat as we move to the
weak scale but  still remains below $m_\mu$\footnote{Recent lattice
calculations seem to prefer such low values}. For example if $\Delta
=1.5$ (corresponding to $v_R$ in the multi-TeV range), we estimate 
$\theta^\ell\simeq \theta_C$.  This
leads to a prediction for the solar mixing angle\cite{smirnov} :
\begin{eqnarray}
sin^2\theta_\odot\simeq 0.32
\end{eqnarray}
The present 3$\sigma$ fit to the solar and KamLand data gives
sin$^2\theta_\odot=0.23-0.38$. It also leads to a prediction for
the $\theta_{13}\simeq \frac{\theta_C}{\sqrt{2}}\simeq 0.15$,
which can be tested in the next round of searches for this
parameter.

In this model, we get $m_e\sim m_0-\frac{9\theta^2_C m^2_s}{\Delta^2
m_\mu}$ which for $m_s\simeq 75$ MeV is about $m_0- 8$ MeV. We can adjust
$m_0$ to get the right value of $m_e$.

The next issue we need to address is the origin of solar mass difference
square. It can arise in a manner which does not affect the quark-lepton
complementarity by adding a gauge singlet $\sigma'$ with $X=2$. This
allows a term of the form 
 $\lambda S_1S_1\sigma'$ with $\lambda\sim 10^{-2}$. This
introduces a $\mu_{11}$ entry which the leads via double seesaw to the
solar mass difference of the right order for $<\sigma'>\simeq
\mu$.

\bigskip
\bigskip

\section{A possible trinification model example}
Another gauge model which can also provide a realization of QLC is
the trinification model based on the gauge group $[SU(3)]^3$\cite{DGG}
which arises from the grand unification group $E_6$. Here we assign the
fermions to
\begin{equation}
\psi_{i} = \psi_R+\psi_L +\psi_c \equiv (3, 3^*, 1)_{Li} + (3^*, 1,
3)_{Li} + (1, 3, 3^*)_{Li}
\label{333ferms}
\end{equation}
Here the subscript (C,L,R) denotes the group under which the multiplet is
a singlet. We then
 add three singlet fermions
\begin{equation}
S_i = (1, 1, 1)_i
\label{333sings}
\end{equation}
For scalars, we use
\begin{equation}
\Phi = \phi_c + \phi_L + \phi_R
\label{333scalars}
\end{equation}
where $\phi_c \sim (1, 3, 3^*) + c.c.$,
$\phi_R \sim (3, 3^*, 1) + c.c.$,
 and $\phi_L \sim (3, 1, 3^*) + c.c.$

Again we impose a $X = f_1 - f_2 - f_3$ symmetry. In addition before
symmetry breaking there is an
$SU(3)_C \times SU(3)_L \times SU(3)_R \times Z_3$
quasi-simple symmetry where the $Z_3$ interchanges the $SU(3)_i$.
Here the $SU(3)_L$ plays a role similar to the generalized weak isospin
of the 331-model\cite{FPP}.
However, it is necessary to respect the fully unified simple
$E_6$ symmetry to obtain the required relation
between quark and lepton couplings.

There are two types of Yukawa couplings possible in this theory: for
quarks the Yukawa responsible giving mass are of the form
$\psi_L\psi_R\phi_c$ whereas for leptons it has the form
$\epsilon^{abc}\epsilon_{pqr}\psi^p_{c,a}\psi^q_{c,b}\phi^r_{c,c}$. At the
$[SU(3)]^3$ level, these two couplings are independent and therefore there
is no connection between quarks and leptons. However, if the theory is
assumed to emerge from an $E_6$ GUT model, the two couplings arise from
the same $[27]^3$ coupling and become equal. In order to derive QLC
relations, we need to assume this. The leptonic part of the Yukawa
interaction can then be written as ($i,j,k$ are generation indices):
 \begin{equation} {\cal L}_{Y} = h_{ij} \psi_{c,i} \psi_{c,j} \phi_c +
f_{ij} \psi_{c,i} S_{Lj} \phi^*_c + \mu_{ij} S_{Li}S_{Lj}
\label{333Yukawa}
\end{equation}
In this 333-model, the VEVs break the $Z_3$ symmetry. The
$<\phi_c>$ has the form
\begin{eqnarray}
<\phi_c>~=~\pmatrix{v_u & 0 & 0\cr 0 & v_d & v_R \cr 0 & v_L & M_3}.
\end{eqnarray}
There are two kinds of vev's in $<\phi>$: the $v_{u,d,L}$ are of order of
the weak scale whereas $v_R$ and $M_3$ are in the multi-TeV range and they
 break $SU(3)_R$. These mass matrices lead to the double seesaw neutrino
mass matrix given above. The rest of the considerations are similar to the
previous model and we do not go into the details here.

This $SU(3)^3 \times Z_3$ model has the advantage
that it is obtainable as a quiver theory\cite{F}
from superstring theory by using the AdS/CFT correspondence
\cite{AdSCFT}.

\section{Discussion}

\bigskip
\bigskip

In this paper, we have presented two possible  gauge models that
realize the quark-lepton complementarity in the 12-sector. An
essential ingredient in our first approach is the 
Pati-Salam $SU(4)_c$ gauge
symmetry which has been suspected in ref.\cite{smirnov} to be a
possible way to obtain the QLC relation. For the first time, we
present explicit models that realize this relation. As we have
noted in the body of the paper, the constraints required to
obtain the QLC relation poses highly nontrivial challenges for
model building. What we show is that it is possible to have a
natural realization under a plausible set of assumptions.
The double seesaw appears to be crucial in our
discussion as is the inverted mass spectrum for neutrinos that is a
consequence of $L_e-L_\mu-L_\tau$ symmetry. The model predicts a value of
$\theta_{13}$ not far below its present upper limit.

\section*{Acknowledgements}

\bigskip
\bigskip

The work of P.H.F. is supported by the US Department
of Energy Grant No DE-FG02-97ER-41036 and that of R. N. M. by the
National Science Foundation Grant No. PHY-0354401. One of us (R. N. M.)
would like to thank S. F. King, H. Minakata and A. Smirnov for comments
and discussions.

\end{document}